\documentclass[twocolumn,pra, floatfix]{revtex4-1} 

\usepackage{graphicx}
\usepackage{amsmath, amssymb, color}

\usepackage{url}

\newcommand{\ket}[1]{\lvert#1\rangle}

\newcommand{\braket}[2]{\langle#1\vert#2\rangle}

\begin{document}
 
\title{Repulsive to attractive interaction quenches of 1D Bose gas in a harmonic trap}

\author{Wladimir Tschischik}
% \author{Roderich Moessner}
\author{Masudul Haque}

\affiliation{Max-Planck-Institut f\"ur Physik komplexer Systeme, N\"othnitzer
Strasse 38, 01187  Dresden, Germany} 

\date{\today}

\begin{abstract}

  We consider quantum quenches of harmonically trapped one-dimensional bosons from repulsive to
  attractive interactions, and the resulting breathing dynamics of the so-called
  `super-Tonks-Girardeau' (sTG) state.  This state is highly excited compared to the ground state of
  the attractive gas, and is the lowest eigenstate where the particles are not bound or clustered.
  We analyze the dynamics from a spectral point of view, identifying the relevant eigenstates of the
  interacting trapped many-body system, and analyzing the nature of these quantum eigenstates.  To
  obtain explicit eigenspectra, we use Hamiltonians with finite-dimensional Hilbert spaces to
  approximate the Lieb-Liniger system.  We employ two very different approximate approaches: an
  expansion in a truncated single-particle harmonic-trap basis and a lattice (Bose-Hubbard) model.
  We show how the breathing frequency, identified with the energy difference between the sTG state
  and another particular eigenstate, varies with interaction.

\end{abstract} 

%\pacs{67.85.-d, 67.85.Hj, 03.75.Kk, 67.85.De}

%% 67.85.-d   Ultracold gases, trapped gases
%% 67.85.De   Dynamic properties of condensates; excitations, and superfluid
    % % flow 
%% 67.85.Jk   Other Bose-Einstein condensation phenomena
%% 67.85.Hj   Bose-Einstein condensates in optical potentials 
%%
%% 03.75.-b   Matter waves 
%% 03.75.Kk   Dynamic properties of condensates; collective and hydrodynamic
    % % excitations, superfluid flow
%% 03.75.Nt   Other Bose-Einstein condensation phenomena
%%
%% 67.10.-j   Quantum fluids: general properties
%% 67.10.Fj   Quantum statistical theory
%%
%% 05.70.-a   Thermodynamics 
%% 05.70.Ln   Nonequilibrium and irreversible thermodynamics
%%
%% 75.10.Pq   Spin chain models
%% 03.65.Xp   Tunneling, traversal time, quantum Zeno dynamics
%% 03.67.Hk   Quantum communication
%% 75.10.Jm   Quantized spin models
%% 75.45.+j   Macroscopic quantum phenomena in magnetic systems

\maketitle

\section{Introduction \label{sec:introduction}}

The traditional focus of many-body quantum physics has been on the ground state and low-energy parts
of the eigenspectrum.  This is well-justified in solid-state systems which are usually in contact
with a thermal bath and typically relax fast to low-energy sectors.  As a result, parts of the
many-body eigenspectra away from the low-energy sector were generally considered to be irrelevant.
% , for much of the history of many-body physics.

In recent years, the perspective has changed dramatically, largely due to new experimental
possibilities with cold atom gases \cite{Cold-atom_reviews_expts}, which have promoted the study of
non-equilibrium situations in \emph{isolated} quantum systems \cite{noneq_reviews}.  In an isolated
situation, energy conservation ensures that a system with an initially high energy will not reach
the low-energy parts of the spectrum; the low-energy sector may thus be unimportant.  
One well-known example is that of `repulsively bound' states of lattice particles
\cite{Winkler-etal_Nature06, bound_clusters_itinerant}.
Eigenstates with particles clustered together have larger energy for repulsive interactions, but
bound states may be unable to decay due to energy conservation.  Eigenstates with bound nature will
determine the dynamics if a system is initially prepared with bound states, even if such eigenstates
are high in energy.

The so-called super-Tonks-Girardeau (sTG) state, which will be the focus of this paper, concerns the
reverse situation of attractive interactions, for bosons confined to one dimension.  Now, clustered
states are lower in energy, and eigenstates without bound states are high up in energy.  Again, in
isolation, once the system is initiated in a `gas-like' state, such high-energy eigenstates
determine the dynamics and the lower-energy cluster states play no role.  In Ref.\
\cite{Naegerl_Science2009}, the lowest gas-like state of a one-dimensional trapped Bose gas was
excited by starting from the ground state of the gas with repulsive interaction and rapidly
switching the sign of the interaction.  For large negative magnitudes, the lowest gas-like state is
called the sTG state.  We will extend the definition slightly and refer to the lowest gas-like state
at any negative interaction as the sTG state.

In this work, we study the properties of the sTG state and the result of a repulsive-to-attractive
interaction quench, in the presence of an explicit harmonic trap.  We provide a \emph{spectral}
description of the trapped system and of the breathing mode dynamics excited by the quench.  

The continuum Lieb-Liniger Hamiltonian has infinite Hilbert space dimension; we address it by two
different approximations with finite Hilbert space dimensions: (1) we approximate it by expressing
the multi-particle basis in terms of a truncated basis of single-particle harmonic oscillator
eigenstates, and (2) we study a tight-binding lattice system (Bose-Hubbard model) with a harmonic
trap.  For these model Hamiltonians we characterize the full energy spectrum of multi-boson systems
as a function of the two-body interaction.  In both systems, there are eigenstates where some or all
particles are (anti-)bound, so that there is a pronounced, roughly linear, dependence of such
eigenenergies on the interaction strength.  Eigenstates without such binding have eigenenergies that
look mostly horizontal when plotted against interaction.  We refer to the first type as `cluster' or
`non-horizontal' states, and the second type as `gas-like' or `horizontal' states.  We identify the
sTG state as the lowest `horizontal' state.

We identify the eigenstates that are excited in a repulsive-to-attractive quench.  In addition to
the sTG state, the important eigenstates are the ground state of the attractive system and the third
horizontal state above the sTG state.  This latter eigenstate is responsible for setting the
breathing mode frequency at large interactions; we will call it the `breathing' eigenstate.  We
present the interaction dependence of the frequency determined by this eigenstate, and explain this
dependence through an examination of the natures of the first few `horizontal' eigenstates.

The breathing mode frequency is twice the trapping frequency, $2\omega_0$, at zero and $\pm\infty$
interactions.  For positive interactions, it is known to fall below this value at intermediate
interactions \cite{our_PRA_2013, KroenkeSchmelcher_PRA2013}.  The sTG state connects to the ground
state of the repulsive case at zero and $\pm\infty$ interactions, and therefore also has breathing
frequency $2\omega_0$ in these limits.  However, at intermediate attractive interactions, the
breathing frequency was predicted \cite{AstrakharchikPRL} using local density approximation
calculations to rise above $2\omega_0$; this was observed in the experiment
\cite{Naegerl_Science2009}.  Our results in this paper give a spectral view of this phenomenon: we
are able to identify the eigenstates in the interacting few-body spectrum whose energy difference
from the sTG state serves as the breathing mode state.  We find that the relevant eigenstate is
different for small and large negative interactions: the third `horizontal' eigenstate plays this
role at large interactions but the ground state plays this role at smaller interactions.

The sTG state was first proposed in Ref.~\cite{AstrakharchikPRL} using quantum Monte Carlo methods.
Combined with a local density approximation for trapped systems, this work predicted the
non-monotonic interaction-dependence of the breathing mode.  Ref.~\cite{TempfliZollnerSchmelcher}
presented multi-configuration time-dependent Hartree calculations for the structure of eigenstates
of the attractive trapped 1D Bose gas.  The experiment of Ref.~\cite{Naegerl_Science2009} created
the sTG state, and measured its breathing mode frequency, using a quench from repulsive to
attractive interactions.  Subsequently, the \emph{homogeneous} sTG gas has been studied using the
Bethe ansatz \cite{Batchelor,ChenBetheAnsatz1,ChenBetheAnsatz2,
  GirardeauAstrakharchikWaveFunctionssTG,KormosMussardoTrombettoni,Caux} and exact diagonalization
\cite{ChenOverlapWithoutTrap}.  The TEBD simulations of Ref.~\cite{Fleischhauer_TEBD} use a weak
trap but do not study trap-generated features like collective modes.  
Analogs of the sTG state in several other systems have been studied theoretically, e.g., in spinor
Bose gases \cite{Girardeau2}, in dipolar Bose gases \cite{sTG_with_dipolar_bosons}, and in fermionic
systems \cite{FermiGases}.

While the homogeneous case has been studied in much greater detail, the experimental realization of
the sTG state is in a harmonic trap and the main diagnostic is the behavior of the breathing mode
excitation, which is only well-defined for a trapped system.  The previous treatments of collective
modes in trapped systems used a local density approximation on the results from equilibrium
calculations with a homogeneous system \cite{AstrakharchikPRL,ChenBetheAnsatz1}.  In contrast, our
work provides a direct study of the breathing mode in the trapped system, both through real-time
evolution after the quench and through a detailed study of the spectrum of the trapped attractive
system.

In Sec.\ \ref{sec:hamiltonian} we introduce the two approximate model Hamiltonians that we use. In
Sec.~\ref{sec:sTGInSpectrum} we describe the full spectrum and the position of the sTG state within
the full spectrum.  In Sec.\ \ref{sec:Quenches} we consider quenches from repulsive to attractive
interactions that preserve the magnitude of the interaction.  We show which eigenstates are excited
in such quenches.  In Sec.~\ref{sec:breathingFrequency} we study the dependence of the breathing
mode frequency on the interaction strength, following both the energy difference between relevant
eigenstates and the frequency extracted from real-time dynamics.  Sec.~\ref{sec:summary} provides a
summary and discussion, and some details are provided in the Appendices.

\section{Model and Hamiltonian \label{sec:hamiltonian}}

We are interested in bosons in one-dimensional traps, interacting through contact interactions,
i.e., the trapped Lieb-Liniger (LL) \cite{LiebLiniger} gas.  The Hamiltonian is
\begin{multline}
 H_{LL} ~=~ \sum_{i=1}^N \left[ 
 -\frac{\hbar^2}{2m} \frac{\partial^2}{\partial x_i^2}
   ~+~\frac{m\omega_0^2}{2}x_i^2 \right]  \\ 
~+~   g\sum_{1\leq i< j \leq N} \delta(x_i - x_j)  \, ,
\label{eq:LL1}
\end{multline}
where $N$ is the number of bosons in the trap, $g$ is the interaction strength, $m$ is the boson
mass, and $\omega_0$ is the trapping frequency.  
%
% For s-wave scattering, $g$ is related to the scattering length $a_{1D}$ as
% $g=-\left(\frac{2\hbar^2}{ma_{1D}}\right)$.
%
Ultracold bosonic atoms behave as a 1D system when the transverse degrees of freedom are frozen out
by tight confinement \cite{Olshanii_PRL1998, Petrovetal_PRL2000}.  The system of Lieb-Liniger bosons
in a harmonic trap has by now been realized in multiple cold-atom laboratories \cite{Naegerl_Science2009,
  trapped_1D-boson_expts}.  

% http://dx.doi.org/10.1016%2Fj.aop.2004.09.010

The continuum LL Hamiltonian above has an infinite-dimensional Hilbert space. In this work, we
address the energy spectrum of the LL Hamiltonian by two different  finite-dimensional
approximations. 

One approach is to represent the many-particle basis in terms of occupations of single-particle
harmonic oscillator eigenstates, with a truncation of the single-particle states (Appendix
\ref{sec:cutoffMdep}).  The resulting Hamiltonian, which we will refer to as the harmonic oscillator
(HO) Hamiltonian, is
\begin{equation}
 H_{HO}^M = \sum_{k=1}^M \left(\frac{1}{2} + k\right) n_k + g
      \sum_{\substack{k,l,\\m,n=1}}^M f_{klmn} a_k^\dagger a_l^\dagger
a_m a_n  \, .
\label{eq:LL2}
\end{equation}
Here $a_k^\dagger$, $a_k$, $n_k={a_k^\dagger}{a_k}$ are bosonic operators for the single-particle
harmonic-oscillator mode $k$, and $f_{klmn}$ is an integral over four harmonic oscillator
eigenfunctions.  In this representation, the kinetic and potential parts of the Hamiltonian are
diagonal, and the interaction part is off-diagonal.
The first $M$ single-particle states are used.  The resulting spectrum contains a finite number,
$\tbinom{M+N-1}{N}$, of eigenstates.  In the limit $M\rightarrow \infty$, the Hamiltonian
\eqref{eq:LL2} is identical to Lieb-Liniger Hamiltonian \eqref{eq:LL1}.  For full numerical
diagonalization, we restrict to $M \le 100$.  The influence of the cutoff $M$ on the spectrum is
discussed in Appendix \ref{sec:cutoffMdep}.  This approach to trapped interacting bosons has been
used, e.g, in Refs.~\cite{HO_diagonalizations_various, Papenbrock_PRA2012}.  

The other approach is to address the physics of the LL gas with a 1D lattice Bose-Hubbard (BH)
Hamiltonian.  This tight binding Hamiltonian describes bosons with on-site interactions and captures
continuum physics for small fillings. The BH Hamiltonian with a  trapping potential is
\begin{equation}
 \begin{split}
    H_{BH} ~=~ &- J \sum_{i=1}^{L-1} \left( b_{i}^\dagger b_{i+1}  + 
    b_{i+1}^\dagger b_{i}  \right) \\
	  &+ \frac{U}{2} \sum_{i=1}^{L}  n_{i}(n_{i}-1)   ~+~
\sum_{i=1}^{L} V(i)n_{i},  
 \end{split}
 \label{eq:BH}
\end{equation}
where $b_{i}$, $b_{i}^\dagger$ are the bosonic operators for site $i$. The Hilbert space is finite
because we use a finite chain of $L$ sites, with $N{\ll}L$.  The parabolic trapping potential
\begin{equation}
V(i) ~=~ \frac{1}{2}k\left( i- \frac{L+1}{2}\right)^2  
\end{equation}
is centered at the midpoint of the chain.  For low enough densities, we can approximate the cosine
dispersion of a lattice particle by a quadratic dispersion; this ``effective mass approximation''
$m^*=\frac{1}{2J}$ ascribes a continuum mass to lattice particles, so that we can relate our
trapping strength $k$ to the trapping frequency of a continuum trapping potential
$\frac{1}{2}m\omega_0^2x^2$:
\begin{equation}
 \omega_0 = \sqrt{2kJ}.
\end{equation}
We will take this to be the definition of the trapping frequency $\omega_0$ on the lattice.

The continuum 1D Bose gas is commonly characterized by the ratio $\gamma$ between interaction and
kinetic energies. For a homogeneous continuum gas, $\gamma = mg/(\hbar^2n)$, where $n$ is the
density. For the dilute BH system $\gamma$ is equal to $\gamma = U/J$ \cite{TGgasOpticalLattice}.
Thus, if we want to compare a low-filling lattice BH gas with a continuum gas having the same
$\gamma$, we would have
\begin{equation}
 U = \frac{mJ}{\hbar^2n} g \, .
\end{equation}
The correspondence is density dependent.  Thus, in the trapped case of interest in this work, where
the density $n$ is not constant, there is no direct correspondence between $g$ and $U$.
Nevertheless, we believe that the comparisons between the spectra and dynamics of the two systems,
which we present in this work, are useful and informative.

In the following, for the BH system, we use $J=\hbar=1$ and therefore measure energy [time] in units
of the tunnel coupling $J$ [inverse tunnel coupling $1/J$].  For the continuum HO Hamiltonian we
measure energy in units of the trapping frequency $\omega_0$, time in units of $T_0= 2\pi/\omega_0$,
and length in units of the harmonic oscillator length $\sqrt{\hbar/m\omega_0}$.

\section{Band structure in spectrum  and location of sTG state 
\label{sec:sTGInSpectrum}}

In this section we describe the energy spectrum for the two finite-dimensional Hamiltonians, the
lattice BH Hamiltonian and the HO Hamiltonian.  In particular, we focus on the location of the sTG
state.  Ref.~\cite{TempfliZollnerSchmelcher} provides similar (but not equivalent) information.

\subsection{ Bose-Hubbard Hamiltonian $H_{BH}$ }

%%%%%%%%%%% FIGURE %%%%%%%%% FIGURE %%%%%%%%%%%%%% FIGURE %%%%%%%%% FIGURE 
\begin{figure}[tb]
\centering
\includegraphics[width=.98\columnwidth]{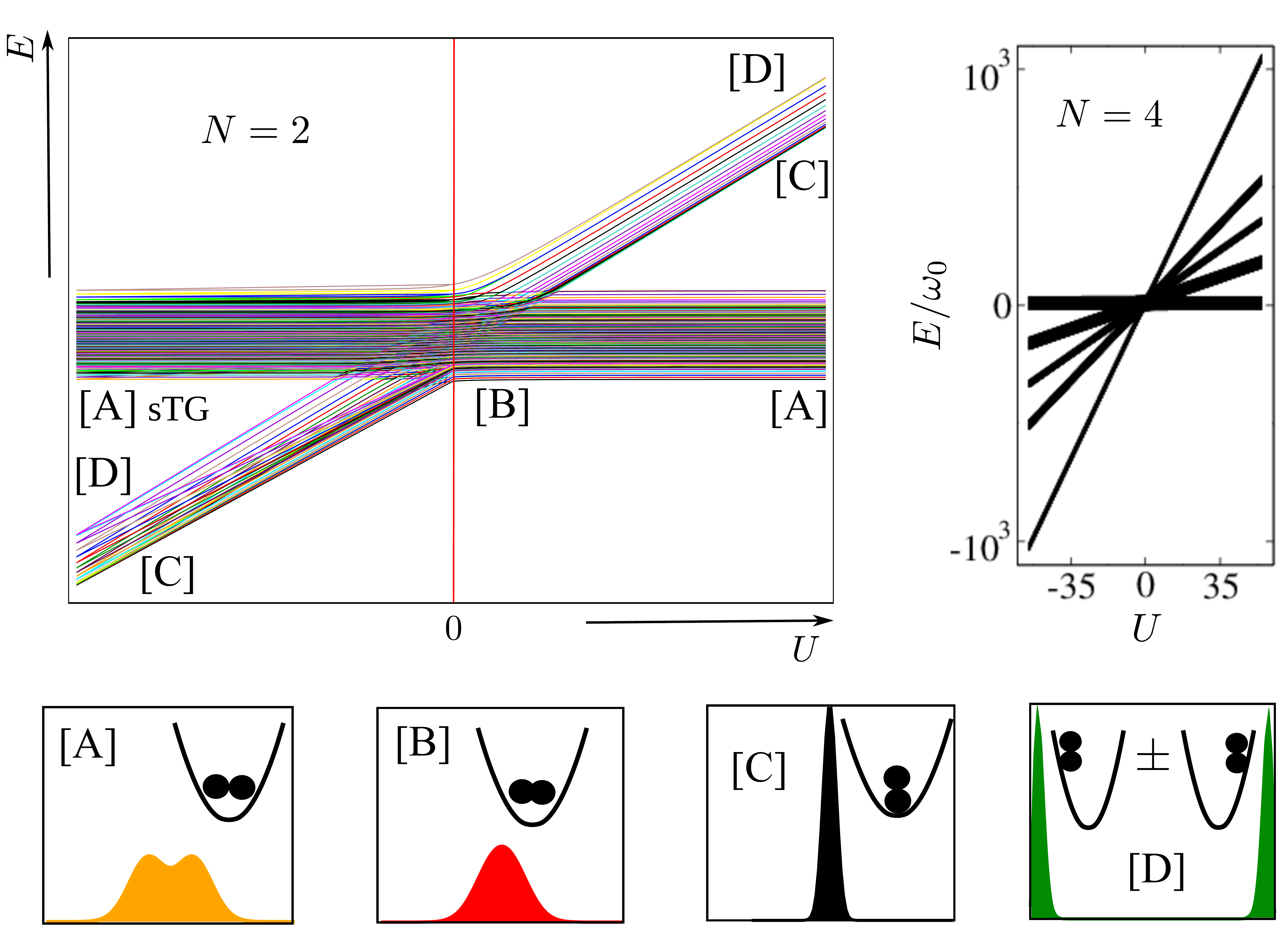}
\caption{\label{fig:sketchSpectrum} 
% 
% (Color online) 
%
Energy spectrum of lattice BH Hamiltonian. 
(Top left) Sketch for $N=2$ bosons.  
The nature of the eigenstates in different regions of the spectra, marked [A] through [D], are shown
in the lower row.  Each panel in the lower row displays eigenstates in two different ways:  using
density profiles (filled curves) and using cartoons showing probable  positions of the two particles. 
(Top right) Spectrum for $N=4$ particles using $L=20$ sites and $k=0.05$. 
}
\end{figure}
%%%%%%%%%%% FIGURE %%%%%%%%% FIGURE %%%%%%%%%%%%%% FIGURE %%%%%%%%% FIGURE 

We begin with the BH Hamiltonian in a harmonic trap. For illustration, Fig.~\ref{fig:sketchSpectrum}
shows the energy spectrum of a BH system of $N = 2$ lattice bosons with a harmonic trap $V(i)$.  At
larger $|U|$, the spectrum separates into two bands.  The horizontal band has states with two
particles separated (gas-like states) and the band with finite slope consists of $L$ eigenstates in
which the boson pair is spatially bound (bound or cluster states).  Cluster states have larger
energy for repulsive interaction.  In contrast, for attractive interactions, the ground state is a
cluster state.

Unlike a Bose-Hubbard chain without a trap, the levels near the bottom of each band are roughly
equally spaced with spacing $\hbar\omega_0$, because the lower parts of each band resembles the
physics of a trapped continuum system, provided the filling is small enough.

For larger number of particles the number of bands increases. Individual bands correspond to states
with different-size clusters, e.g., clusters of $2$, $3$,... , $N$ bound bosons and possible
combinations. The right panel of Fig.~\ref{fig:sketchSpectrum} shows the spectrum for $N=4$ bosons with
five bands.

At small attractive interactions, bands with cluster structure cross with gas-like states, making
the identification and behavior of gas-like states more complicated.  At large $|U|$ the bands are
completely separated.  This occurs when the interaction energy sets the largest energy scale, so
that the effect of negative $U$ wins over even the largest trap energy that can be gained by the
particles by sitting at the edges of the lattice (states of type [D] in Fig.~\ref{fig:sketchSpectrum}, at the top of the cluster band).

\subsection{ Harmonic-Oscillator Hamiltonian $H_{HO}$ 
  \label{sec:continuumLattice}}

%%%%%%%%%%% FIGURE %%%%%%%%% FIGURE %%%%%%%%%%%%%% FIGURE %%%%%%%%% FIGURE 
\begin{figure}[tb]
\centering
\includegraphics[width=.98\columnwidth]{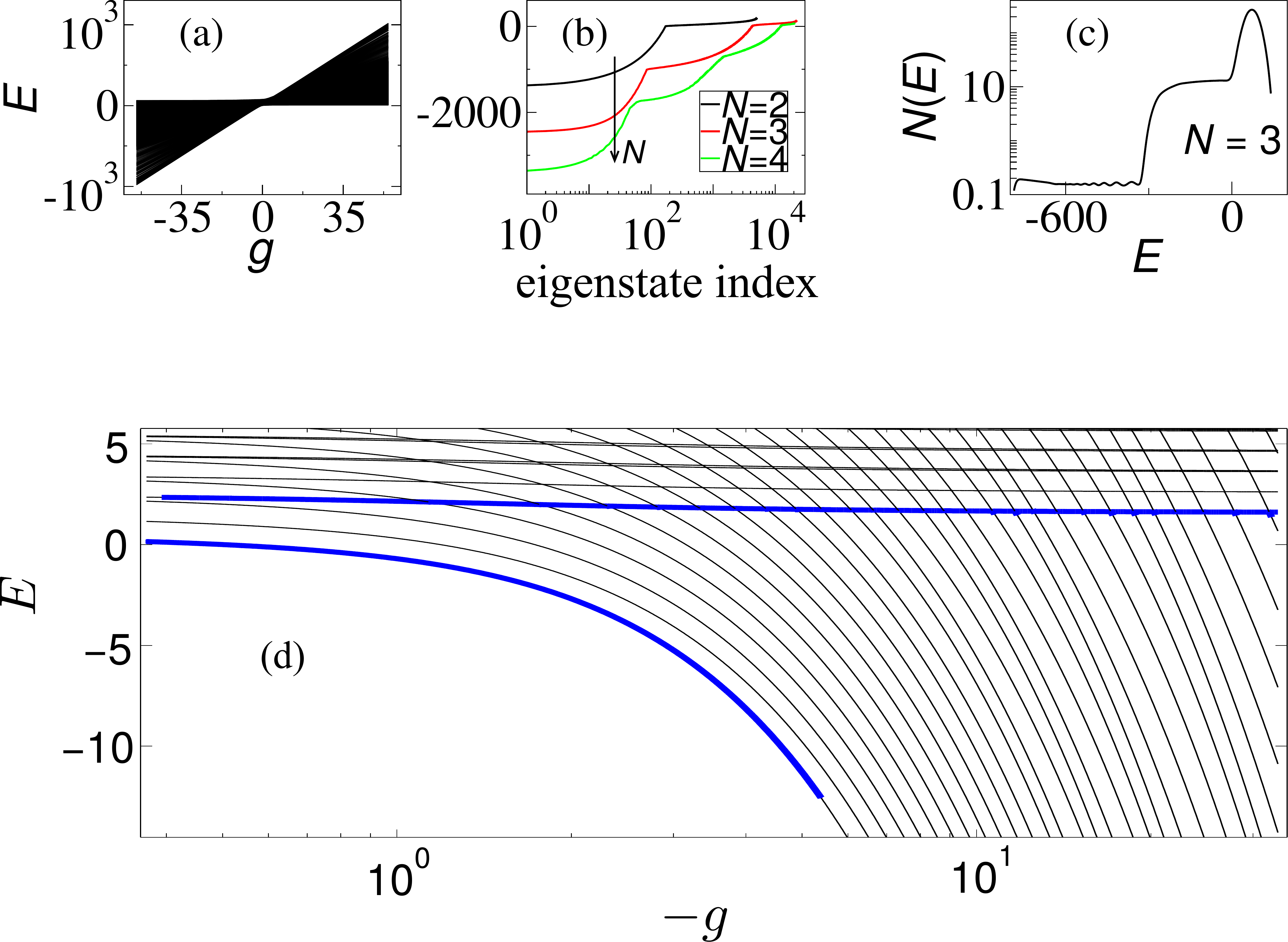}
\caption{ \label{fig:SpectrumContinuum}
%
% (Color online) 
%
(a) Energy spectrum of continuum HO Hamiltonian for $N=4$ using $M=20$
  orbitals.  In contrast to BH spectrum, the HO spectrum shows no clear separation of bands for
  large interaction $|g|$.  (b) Eigenenergies of the HO Hamiltonian plotted in increasing order at
  $g = -165$, for $\{N,M\}=\{2,100\}; \{3,50\}; \{4,25\}$. Note logarithmic scale for the
  eigenenergy index.  (c) Density of states for $N=3, M=50, g=-54$.  Eigenenergies and density of
  states both show plateaus, corresponding loosely to the band structure in the  BH spectrum.  (d)
  Energy spectrum of HO Hamiltonian for $N=2, M=20$. Thick lines indicate states having largest
  overlap with the ground state for $g>0$ after $g \rightarrow -g$ quench. STG state is the lowest
  horizontal state.  }
\end{figure}
%%%%%%%%%%% FIGURE %%%%%%%%% FIGURE %%%%%%%%%%%%%% FIGURE %%%%%%%%% FIGURE

Does the band structure of the BH lattice system, described above, survive in some sense in the
continuum case?  This question is examined here by consideration of our second finite-dimensional
Hamiltonian, the HO Hamiltonian \eqref{eq:LL2}.

In Fig.~\ref{fig:SpectrumContinuum}(a) we plot the spectrum for the HO Hamiltonian for $N=4$ bosons
in a trap using $M=20$ single-particle orbitals. This could be compared to the BH energy spectrum
for $N=4$ shown in Fig.~\ref{fig:sketchSpectrum}; the dimension of the Hilbert space is the same in
the two cases. The clear band structure seen in the spectrum of the BH model is not present.
However a signature of the band structure is visible in the density of eigenenergies at large
$|g|$. In Fig.~\ref{fig:SpectrumContinuum}(b) we plot the eigenenergies for large negative $g$, in
ascending order, against the index of ordering.  One can see clearly a plateau structure.  (Note
logarithmic scale for the eigenvalue index.)  Energies of cluster eigenstates strongly depend on the
cutoff $M$; thus for small $M$ the energies of the non-horizontal states are not reliable
(Appendix~\ref{sec:cutoffMdep}).  
%
% For small $M$ the density of states at low energy is small as is seen in
% Fig~\ref{fig:SpectrumContinuum}(c) for $N=3$.

The spectrum of the HO Hamiltonian shows no clear band separation, nevertheless it is possible to
identify the sTG state in the spectrum. In Fig.~\ref{fig:SpectrumContinuum}(d) the sTG state is
visible as the lowest horizontal state, shown with a thicker line.  
(There is some ambiguity in the close vicinity of each level crossing.)

\section{Interaction Quenches to attractive regime \label{sec:Quenches}}

%%%%%%%%%%% FIGURE %%%%%%%%% FIGURE %%%%%%%%%%%%%% FIGURE %%%%%%%%% FIGURE
\begin{figure}[tb]
\centering
\includegraphics[width=.98\columnwidth]{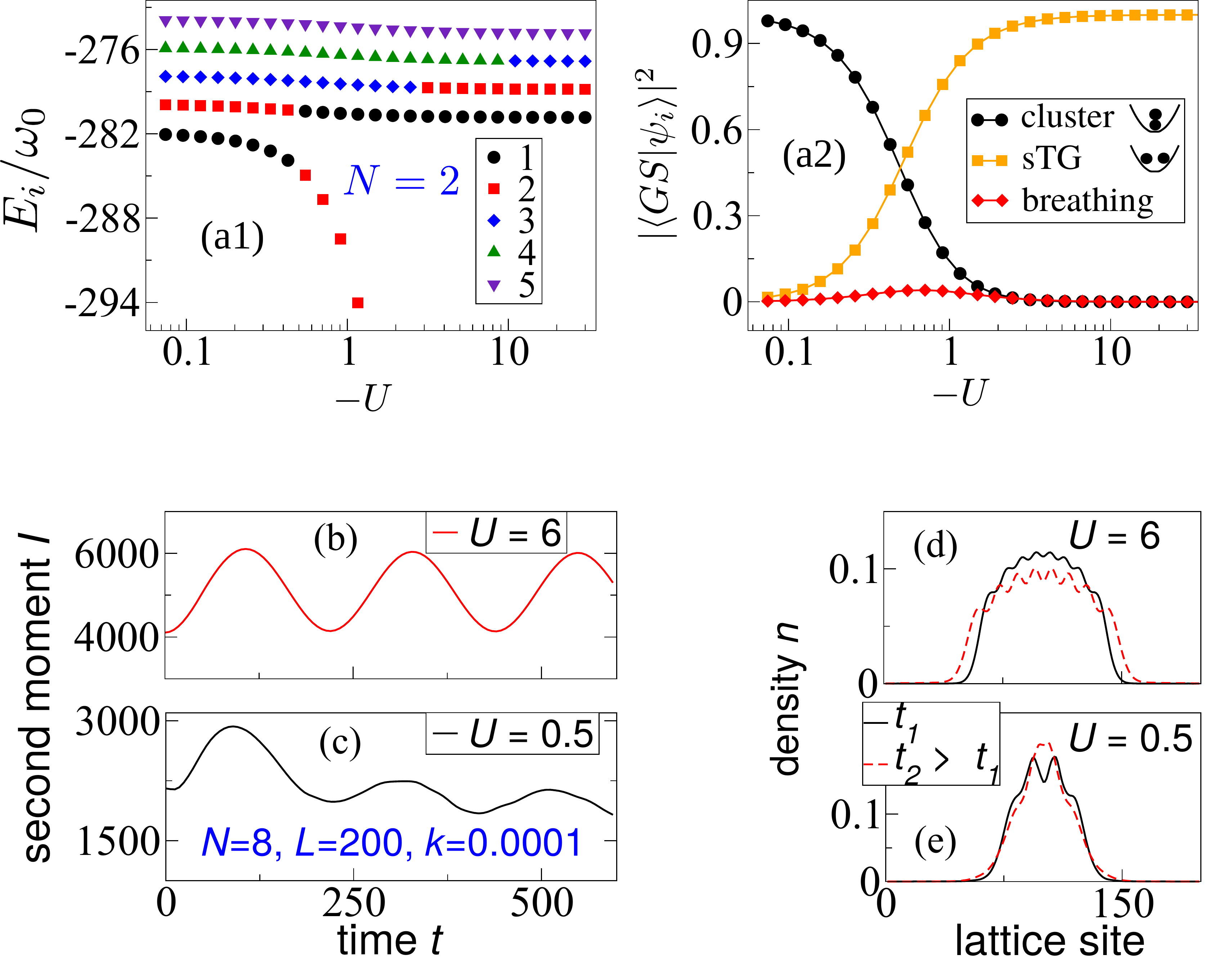}
\caption{    \label{fig:UquenchN6L37}
%
% (Color online) 
%
  $U{\to}-U$ quenches in the lattice BH case. (a1,a2) Data for $N=2$, $L=60$, $k=0.001$.  (a1)
  Energies of the five eigenstates of the post-quench Hamiltonian having largest overlaps with the
  initial groud state.  (a2) Overlap magnitudes; only the largest three are shown.  Outside the
  band-crossing region the overlap with sTG state is almost unity.  
  (b,c) Real-time breathing dynamics after $U{\to}-U$ quenches, computed using TEBD; $N=8$, $L=200$,
  $k=0.0001$.  The dynamics of the cloud size, quantified using the second moment of density $n$, is
  shown.  (d,e) Density distributions at some times during the post-quench dynamics. }
\end{figure}
%%%%%%%%%%% FIGURE %%%%%%%%% FIGURE %%%%%%%%%%%%%% FIGURE %%%%%%%%% FIGURE

In this section we study temporal dynamics induced by repulsive to attractive interaction
quenches. We start from the ground state $\ket{GS}$ of the trapped system with repulsive
interaction.  The sign of the interaction is then switched while keeping the same magnitude, i.e,
$U\to-U$ for the BH Hamiltonian and $g\to-g$ for the HO Hamiltonian.  We identify relevant
eigenstates by looking at overlaps $\left|\braket{GS}{\psi_i}\right|^2$ between the initial state
$\ket{GS}$ and eigenstates $\ket{\psi_i}$ of the post-quench Hamiltonian.
 
We start with the lattice (BH) case.  The ground state $\ket{GS}$ at positive $U$ and its breathing
mode excitation has been detailed in Ref.\ \cite{our_PRA_2013}. The density profile is Gaussian
for small $U$; in the large-$U$ (fermionized) regime it has the characteristic free-fermion shape
with $N$ peaks.

% We begin with repulsive bosonic gas on a lattice. The ground state $|GS\rangle$ of the repulsive BH
% Hamiltonian has a Gaussian density profile and an energy $\frac{1}{2} N \omega_0$ above the band end
% at $-2NJ$ for noninteracting bosons. With increasing $U$ energy increases by
% $\frac{1}{2}N(N-1)\omega_0$ toward hard-core bosons (fermionized) limit for infinite $U$. In this
% limit hard-core bosons behave like noninteracting spinless fermions. The density profile has $N$
% characteristic peaks. Small quenches of the on-site interaction or trapping strength excite the
% breathing mode, time-periodical contraction and expansion of the bosonic cloud. The breathing mode
% frequency is $\Omega_{U=0}=2\omega_0$ at non-/infinity-interacting limits and has a minimum
% $\Omega^{min}_{U>0}= \sqrt{3}\omega_0$ for large $N$ \cite{BreathingModeBH}.

If Fig.~\ref{fig:UquenchN6L37}(a), we identify the $U<0$ eigenstates which have the largest overlaps
with the initial state $\ket{GS}$.  Fig.~\ref{fig:UquenchN6L37}(a1) shows the energies of the five
eigenstates with largest overlap.  The three most important eigenstates are the ground state (which
is a `cluster' state at large $U$), the lowest `horizontal' eigenstate (sTG state), and the third
`horizontal' eigenstate higher in energy from the sTG state.  We show the overlap magnitudes with
these three eigenstates in Fig.~\ref{fig:UquenchN6L37}(a2) as a function of $U$.

At large negative $U$, the ground state is a cluster state, whose structure is very different from
the initial state $\ket{GS}$.  In contrast, the sTG state is close (identical) to $\ket{GS}$ for
large (infinite) $|U|$.  In Fig.~\ref{fig:UquenchN6L37}(a2), this is visible through the large
overlap with the sTG state and vanishing overlap with the cluster state at large $|U|$.  This large
overlap at large interactions has also been pointed out for the trap-free system in
Refs.~\cite{ChenOverlapWithoutTrap, ChenBetheAnsatz1}.  Around $U=0$, the state $\ket{GS}$ at
positive $U>0$ is adiabatically connected to the ground state at $U<0$; hence the overlap with the
final ground state is seen in Fig.~\ref{fig:UquenchN6L37}(a) to be large at small $|U|$.
In the region of the band crossing (small $|U|$) there are more than three states having substantial
($\ge 1\%$) overlap with initial state, although only the three with largest overlaps are shown in
Fig.~\ref{fig:UquenchN6L37}(a2).

In the BH system the band crossing part of the spectrum is qualitatively different from the part of
the spectrum with separated bands.  In the band-crossing region (small $|U|$), non-equilibrium dynamics
following the quench will not be a simple breathing, since many states (hence many frequencies) are
excited.  In contrast, in the band-separated region (large $|U|$), we expect clean breathing
dynamics.  This is seen in the temporal dynamics pictures of Fig.~\ref{fig:UquenchN6L37}(b,c),
where we show the evolution of the second moment of the density (site occupancy) distribution $I =
\sum_i n_i (i-\frac{L+1}{2})^2$, which measures the width of the bosonic cloud.
The time evolution data is  shown for a larger system, computed using Time-Evolving Block
Decimation (TEBD) \cite{TEBD}.  The sizes shown here are too
large for us to obtain eigenstate overlap data, but the predicted feature is clearly seen: clean
breathing mode for large $|U|$ and multi-frequency dynamics for smaller $|U|$ in the band crossing
region.

%%%%%%%%%%% FIGURE %%%%%%%%% FIGURE %%%%%%%%%%%%%% FIGURE %%%%%%%%% FIGURE
\begin{figure}[bt]
\centering
\includegraphics[width=0.98\columnwidth]{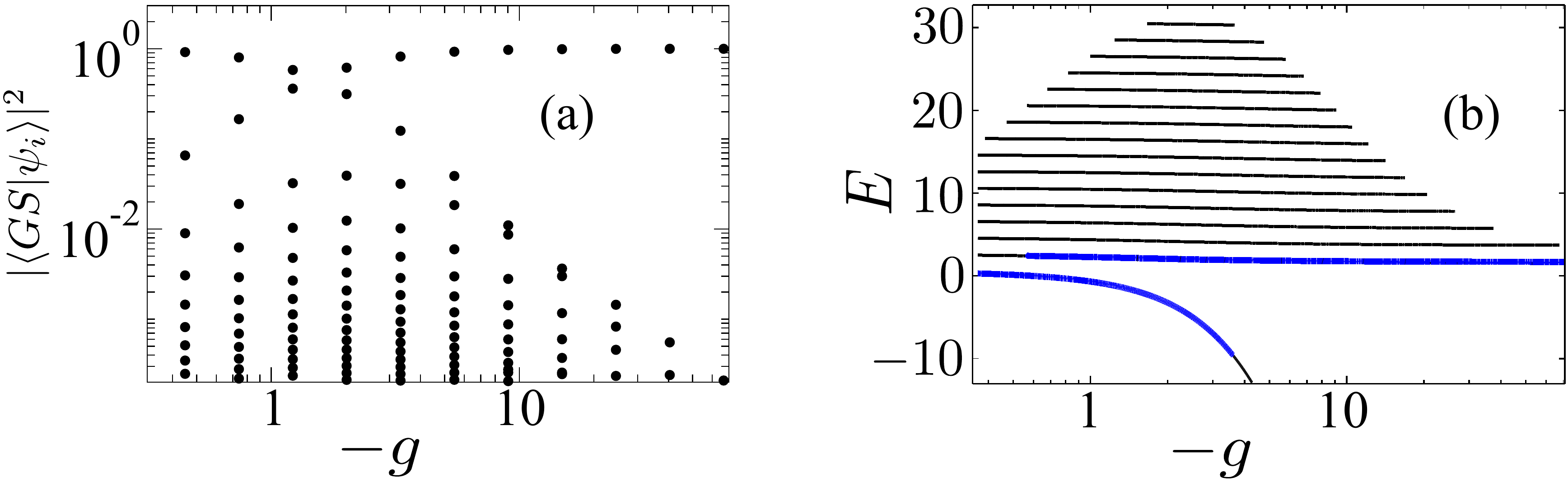}
\caption{  \label{fig:OverlapSpectrumN2M100_02}
 $g\to-g$ quenches in the HO Hamiltonian, $N=2$, $M=100$. 
(a) Overlap of the initial ground state with eigenstates of the
post-quench HO Hamiltonian.
(b) Energies of the states having the overlap larger than $2\times10^{-4}$ at 
different $g$. Thick lines indicate states with the largest overlap.
}
\end{figure}
%%%%%%%%%%% FIGURE %%%%%%%%% FIGURE %%%%%%%%%%%%%% FIGURE %%%%%%%%% FIGURE

In the HO system (Fig.~\ref{fig:OverlapSpectrumN2M100_02}), the ``bands'' are never completely
separated, however the overall situation for $g\to-g$ quenches is similar, at least for $N=2$.  The
overlap is largest with the ground state at smaller $|g|$, and with the sTG state at larger $|g|$.
Analogous to the BH model in the band-crossing region, for moderate $|g|$ a few other ``horizontal''
eigenstates are excited in addition to the lowest horizontal (sTG) eigenstate.

\section{Breathing mode of sTG state \label{sec:breathingFrequency}}

\subsection{Interaction dependence}

Collective modes in a trapped system are generally associated with low-lying eigenstates.  For the
repulsive 1D trapped Bose gas, the eigenstate associated with the breathing mode and its excitation
with respect to the ground state (which is the breathing frequency) has been detailed recently, for
both lattice \cite{our_PRA_2013} and continuum \cite{KroenkeSchmelcher_PRA2013} cases.  The lowest
excited state, at excitation energy around the trapping frequency, has odd spatial parity and
corresponds to dipole oscillations (Kohn mode).  There are two excited states with energy near twice
the trapping frequency, one of them interaction-independent and the other deviating from $2\omega_0$
at finite interactions.  This state with constant excitation energy is part of the equally spaced
sequence of states related to the Kohn (dipole) mode \cite{KohnMode}.  The interaction-dependent
energy level corresponds to the breathing mode.  For large particle number the excitation energy of
the breathing-related eigenstate at intermediate interactions approaches the mean-field prediction
for the breathing frequency: $\sqrt{3}\omega_0 < 2\omega_0$ \cite{our_PRA_2013,
  KroenkeSchmelcher_PRA2013}.

For attractive interactions, the analogous picture for excitations over the sTG state is as follows.
There is a dipole-mode state at energy $\omega_0$ above the sTG state, and two states at energy near
$2\omega_0$.  The breathing-mode state is again the one whose excitation energy is
interaction-dependent, but in this case, this excitation energy is \emph{larger} than $2\omega_0$.
The breathing mode state is thus the \emph{third} gas-like state above the sTG state and not the
\emph{second} one above the sTG state.  In addition, the situation can be complicated by the
presence of many `non-horizontal' states crossing through this energy region.

%%%%%%%%%%% FIGURE %%%%%%%%% FIGURE %%%%%%%%%%%%%% FIGURE %%%%%%%%% FIGURE
\begin{figure}[tb]
\centering
\includegraphics[width=.98\columnwidth]{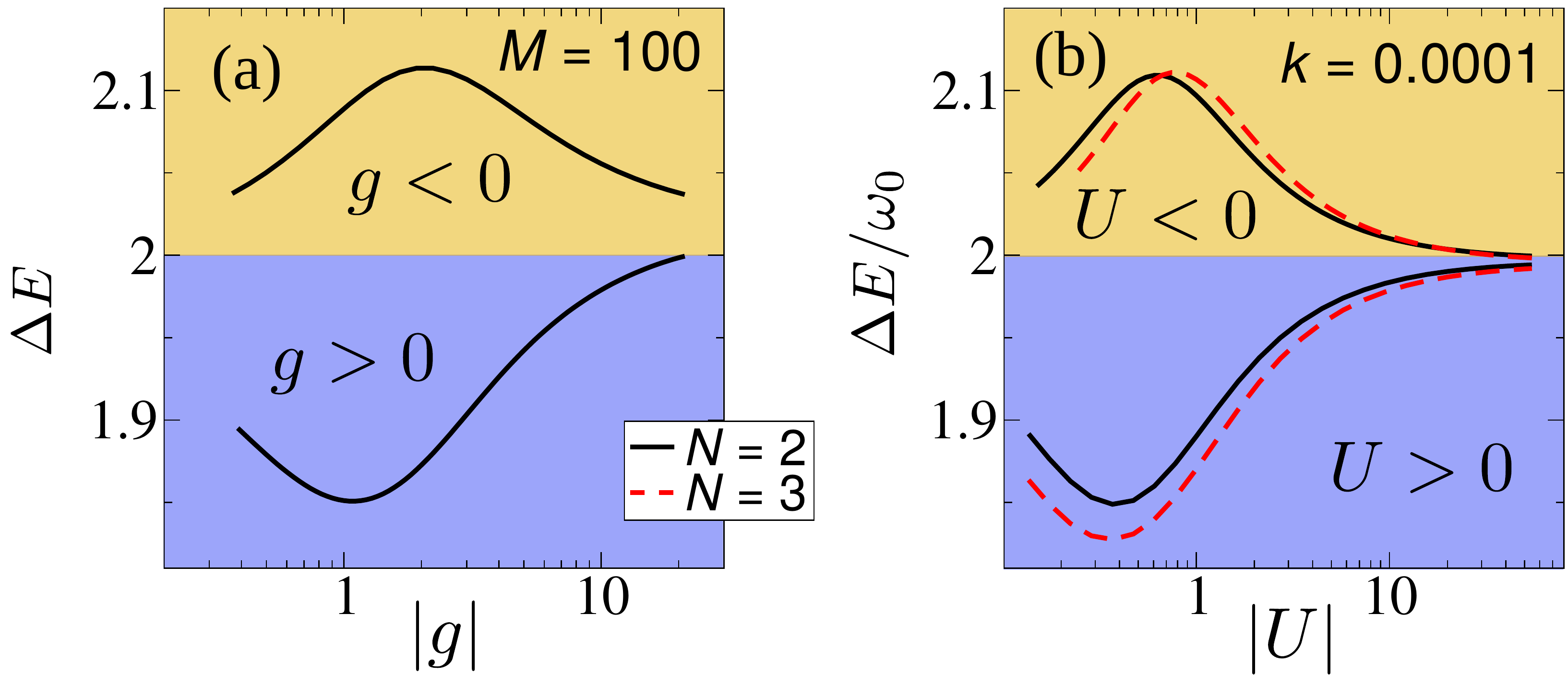}
\caption{  \label{fig:DE_N23_02}
%
% (Color online) 
%
Excitation energy of the breathing mode state above 
the sTG state (the ground state) for attractive (repulsive) interaction. 
The sTG state and the breathing mode state are found by looking at the largest 
overlaps for the interaction quenches from repulsive to attractive regimes.
(a) In the HO spectrum for $N=2$, $M=100$. (b) In the BH spectrum for 
$N=2$, $L=250$ and $N=3$, $L=106$ and $k=0.0001$.
} 
\end{figure}
%%%%%%%%%%% FIGURE %%%%%%%%% FIGURE %%%%%%%%%%%%%% FIGURE %%%%%%%%% FIGURE

As detailed in the previous section, repulsive to attractive quenches at large $|U|$ or $|g|$
populate mainly the sTG state, but also excite breathing modes on top of the sTG state by populating
the relevant breathing mode eigenstate.  Thus, looking at the energy difference between the two
gas-like states with highest overlaps, we can identify the excitation energy of the breathing mode
eigenstate at various interaction strengths.  In Fig.~\ref{fig:DE_N23_02} we show the excitation
energy of the breathing mode state $\Delta{E}$ relative to the sTG state.  The excitation energy
(breathing mode frequency) is larger than $2\omega_0$ and has a maximum at some finite value of the
interaction.  As in the repulsive case \cite{our_PRA_2013, KroenkeSchmelcher_PRA2013}, the breathing
mode frequency is equal to $2\omega_0$ for zero and infinite interactions (with possible deviations
in the BH case due to finite lattice filling \cite{our_PRA_2013}).  For comparison, we also show the
breathing mode frequency for the repulsive case, for both the lattice BH Hamiltonian and the
continuum HO Hamiltonian.

%%%%%%%%%%% FIGURE %%%%%%%%% FIGURE %%%%%%%%%%%%%% FIGURE %%%%%%%%% FIGURE
\begin{figure}[tb]
\centering
\includegraphics[width=1\columnwidth]{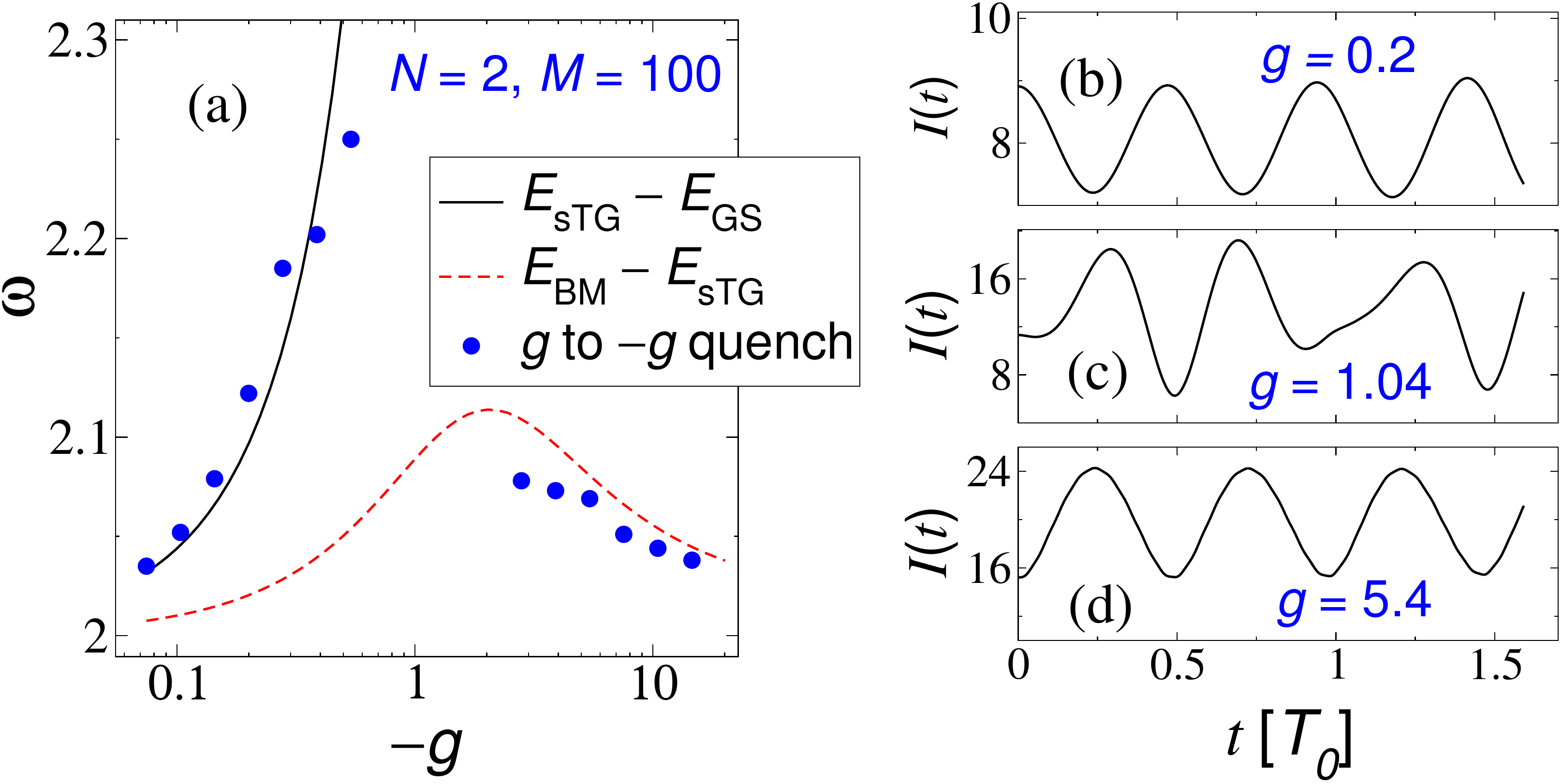}
\caption{ \label{fig:BMF_from_time_evol}
%
%(Color online)
%
HO Hamiltonian with $N=2$ particles.  (a) Frequencies extracted from time evolution of the cloud
width $I(t)$ after $g\to-g$ quench (dots), compared with the energy difference between sTG and
ground states (solid line), and the energy difference between sTG and breathing eigenstate
(dashed, non-monotonic line).  (b-d) Dynamics of the cloud radius after  $g\to-g$ quench.  At
intermediate $g$, the dynamics clearly involves multiple frequencies.  
}
\end{figure}
%%%%%%%%%%% FIGURE %%%%%%%%% FIGURE %%%%%%%%%%%%%% FIGURE %%%%%%%%% FIGURE

The excitation energy of the `breathing mode eigenstate' (third horizontal eigenstate above the sTG
state) corresponds only at larger interactions to the breathing mode frequency obtained through time
evolution.  There are complications at small interactions, because the non-horizontal ground state
plays an important role (has large overlap) in repulsive to attractive quenches.  This is shown in
Fig.~\ref{fig:BMF_from_time_evol}, where the frequency of oscillations of the cloud size $I(t)$
after the quench is compared with energy differences in the spectrum.  At larger interactions, the
frequency matches the energy difference between the sTG state and the third horizontal state above
it.  At smaller interactions, however, the frequency follows much more closely the energy difference
between the sTG state and the ground state.  This is because for small positive to small negative
interactions the dominant eigenstates are the sTG state and the ground state (previous section).
There is an intermediate interaction range where the dynamics clearly involves multiple frequencies
and it is difficult or impossible to assign a single frequency to the evolution of  $I(t)$.

\subsection{Structure of `horizontal' eigenstates}

%%%%%%%%%%% FIGURE %%%%%%%%% FIGURE %%%%%%%%%%%%%% FIGURE %%%%%%%%% FIGURE
\begin{figure}[tb]
\centering
\includegraphics[width=1\columnwidth]{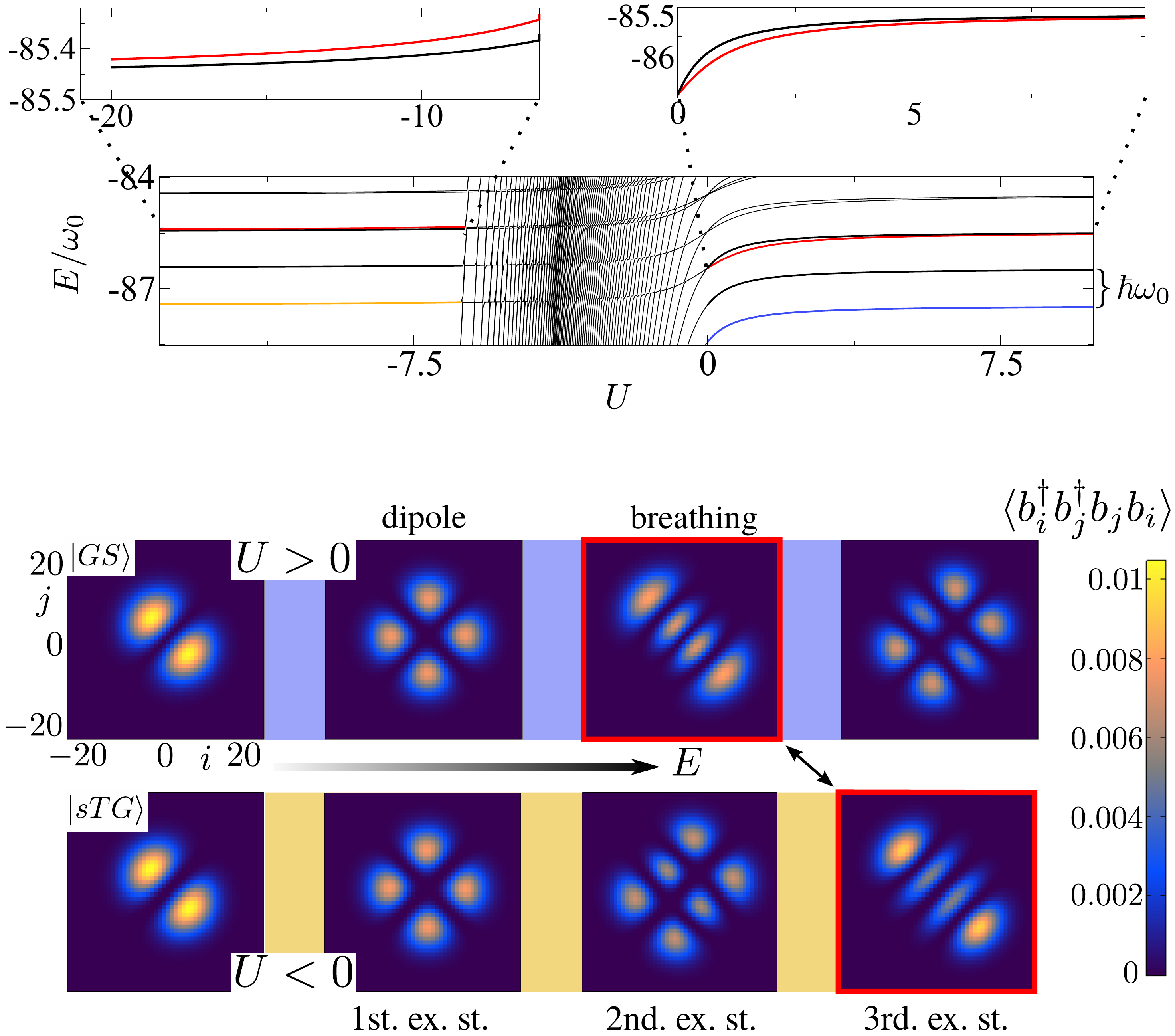}
\caption{ \label{fig:denDenCorrN2L100U10_0_001}
%(Color online) 
%
  (Top) Relevant part of the energy spectrum of BH Hamiltonian for $N=2, L=100, k=0.001$.  (Bottom)
  Density-density correlation $\langle b_i^\dagger b_j^\dagger b_j b_i \rangle =
  \langle{n_i}{n_j}\rangle - n_i \delta_{ij}$ for $N=2$, shown as a density plot in the $(i,j)$
  plane.  Here $L=100, k=0.001$ and $|U|=10$.  Upper panels show  four lowest-energy eigenstates
  for $U>0$.  Lower panels show four lowest gas-like states for $U<0$.  The breathing mode
  eigenstate is the third state on the  $U>0$ side and the fourth gas-like state on the  $U<0$ side.  
}
\end{figure}
%%%%%%%%%%% FIGURE %%%%%%%%% FIGURE %%%%%%%%%%%%%% FIGURE %%%%%%%%% FIGURE

The top panel of Fig.~\ref{fig:denDenCorrN2L100U10_0_001} shows the relevant part of the BH spectrum
for $N=2$ particles.  The second and third horizontal states above the sTG both have around
$2\omega_0$ higher energy than sTG state. As opposed to the repulsive case where the breathing mode
is associated with the second excited state and the third excited state is the Kohn-related
eigenstate \cite{our_PRA_2013}, in the sTG case the breathing mode is associated with the
\emph{third} horizontal excited state while the \emph{second} horizontal excited state is
Kohn-related and has approximately interaction-independent energy compared to the sTG state.

The structure of eigenstates can be visualized through the \textit{density-density correlation}
(DDC) $\langle b_i^\dagger b_j^\dagger b_j b_i \rangle = \langle {n_i}{n_j}\rangle - n_i
\delta_{ij}$.  Fig.~\ref{fig:denDenCorrN2L100U10_0_001} shows plots of $\langle b_i^\dagger
b_j^\dagger b_j b_i \rangle$ for two particles at large interaction, outside the band crossing
region (BH Hamiltonian).  The upper row of four panels show DDC for the four lowest eigenstates of
repulsive ($U>0$) bosons.  The first excited state is associated with the dipole mode and the second
with the breathing mode.  The bottom row shows the sTG state and the three eigenstates above it; we
see the same structures except that the second and third excited state structures are switched in
the two cases.

The dipole mode corresponds to center-of-mass motion, i.e. oscillation of bosonic cloud around a
potential minimum.  The associated eigenstate has greatest intensities at the configurations where
the center of the pair is displaced, specifically, configurations with one particle is at the trap
center and the other displaced.  There are four such configurations in $(i,j)$ plane, and
accordingly four spots in the DDC plot.  During the breathing motion center-of-mass is conserved,
particles move toward and away from each other.  The corresponding eigenstate is dominated by
configurations where the two particles are symmetrically placed around the trap center, but at a
different distance from each other than in the ground ($U>0$) or sTG ($U<0$) state.  The DDC of the
third (second) excited state for $U>0$ ($U<0$) is not obvious to interpret.  This state is part of
the ``Kohn tower'' --- the equally spaced tower of states associated with the many-body ladder
operator that describes dipole oscillations \cite{KohnMode}.

To explain heuristically why the breathing-mode eigenstate is lower (higher) than the Kohn-related
mode for $U>0$ ($U<0$), we note from the DDC plots that the breathing-related eigenstate is marked
by larger distances between the particles.  For $U>0$ ($U<0$), this leads to lower (higher)
interaction energy.  This gives an intuitive explanation of why the breathing frequency for the
$U>0$ ground state is smaller than $2\omega_0$ while that of the $U<0$ sTG state is larger than
$2\omega_0$, as we have shown in Fig.~\ref{fig:DE_N23_02}.  
%
% Note that $2\omega_0$ is the breathing frequency for the non-interacting case.

\section{Summary and Discussion \label{sec:summary}}

We have presented a study of the dynamics involved in exciting the sTG state in one-dimensional
trapped bosons using quantum quenches from repulsive to attractive contact interactions.  We have
focused on a many-body spectral description, using exact numerical diagonalization of models with
finite Hilbert space to examine the spectrum, the structure of eigenstates, and the overlaps of
various eigenstates with the pre-quench state.  We show that the eigenstate responsible for the
breathing is the third horizontal state above the sTG state, which has a maximum above $2\omega_0$
(rather than a minimum below $2\omega_0$) at intermediate negative interactions.  This perspective
on the breathing mode frequency, based on the eigenspectrum, complements previous studies using
local density approximations and sum rules \cite{AstrakharchikPRL,ChenBetheAnsatz1}, and
experimental measurements of the breathing mode frequency \cite{Naegerl_Science2009}.  Examining the
structure of the eigenstates using the density-density correlations, we arrive at an intuitive
argument for the direction of the shift of the breathing frequency from $2\omega_0$.  Examining
explicitly the time evolution of the cloud size, we have found that at small negative interactions
the breathing frequency corresponds to a different energy difference --- the difference between the
lowest horizontal state and the ground state of the system (Fig.\ \ref{fig:BMF_from_time_evol}).

A limitation of the present study is that the effects of finite $M$ (HO Hamiltonian) or finite $L$
(BH Hamiltonian) become severe at larger particle number.  In the HO case, finite $M$ effects affect
the cluster states rather drastically (Appendix \ref{sec:cutoffMdep}).  In the BH case, finite $L$
limitations lead to configurations where the filling is not $\ll1$ at the center of the trap, in
which case the lattice results do not approximate continuum physics well.
Most of our spectral results are obtained from full numerical diagonalization.  Sparse matrix
methods, which would have allowed access to larger $M$ or $L$, are not well-suited to most of the
calculations because the relevant energies are not at the bottom of the spectrum, and because the
density of states can be large due to the presence of many `horizontal' and `non-horizontal' states
in the same spectral region.
Thus, the overlap data in Figs.\ \ref{fig:UquenchN6L37} and \ref{fig:OverlapSpectrumN2M100_02}, and
the excitation energies in Figs.\ \ref{fig:DE_N23_02} and \ref{fig:BMF_from_time_evol}, are shown
for $N=2$ or $N=3$ particles.  For larger $N$, we observe similar qualitative features, but the
corresponding overlaps and energies are generally not quantitatively reliable due to the finiteness
effects mentioned above.  While our particle numbers (2 or 3) are small, the numbers in the
experiment \cite{Naegerl_Science2009} (around 20) are far too small to be justifiably approximated by
mean field theories or the thermodynamic limit or local density approximations.  Our small-system
study of the breathing mode thus provides a valuable counterpoint to previous infinite-number
studies.

The present work raises a number of open questions.  First, we have found the breathing mode
frequency at small and large interactions to be related to different eigenstates.  This raises the
question of which eigenstates are responsible for the breathing mode frequencies in the experiment
\cite{Naegerl_Science2009}.  Second, in both our models, we see an intermediate interaction regime
where we have multi-frequency breathing oscillations; an open question is whether a similar effect
might be visible in a higher-resolution experiment.  Third, we have focused on breathing motion
excited by interaction quenches excited by repulsive to attractive quenches.  If the dynamics is
excited by small interaction quenches or by small quenches of the trapping strength, one might
imagine a different set of eigenstate overlaps, which may lead to differences in the size dynamics
or even frequency.  Finally, the collapse of the breathing mode to zero frequencies for small
interactions found in the local density approximation calculations of Ref.\ \cite{AstrakharchikPRL}
remains a mystery; it is unclear whether this should be identified with the interaction regime where
we see multi-frequency dynamics, or with the interaction regime where the mixing with the ground
state is responsible for the breathing dynamics of the sTG state.

% --- ----

% Praise the spectral perspective for dynamics.  Cite various, e.g., our ladder paper.  

% cite papers and reviews on breathing modes 

% --- ----

% \textcolor{red}{** Further comparison with previous work?  **}

% --- ----

% Part of our study is similar in spirit to Ref.~\cite{TempfliZollnerSchmelcher}; they did not focus
% on collective modes or dynamics after a quench. 

% --------- ---------

\begin{acknowledgments}

  We thank A.~Eckardt, A.~M.~L\"auchli, A.~Lazarides, R.~Moessner, H.-C.~N\"agerl, and D.~Vorberg
  for useful discussions.  We also thank R.~Moessner for collaboration on related work
  \cite{our_PRA_2013}.

\end{acknowledgments}

\appendix

\section{Expansion in free harmonic oscillator basis \label{sec:cutoffMdep}}

%%%%%%%%%%% FIGURE %%%%%%%%% FIGURE %%%%%%%%%%%%%% FIGURE %%%%%%%%% FIGURE
\begin{figure}[tb]
\centering
\includegraphics[width=1\columnwidth]{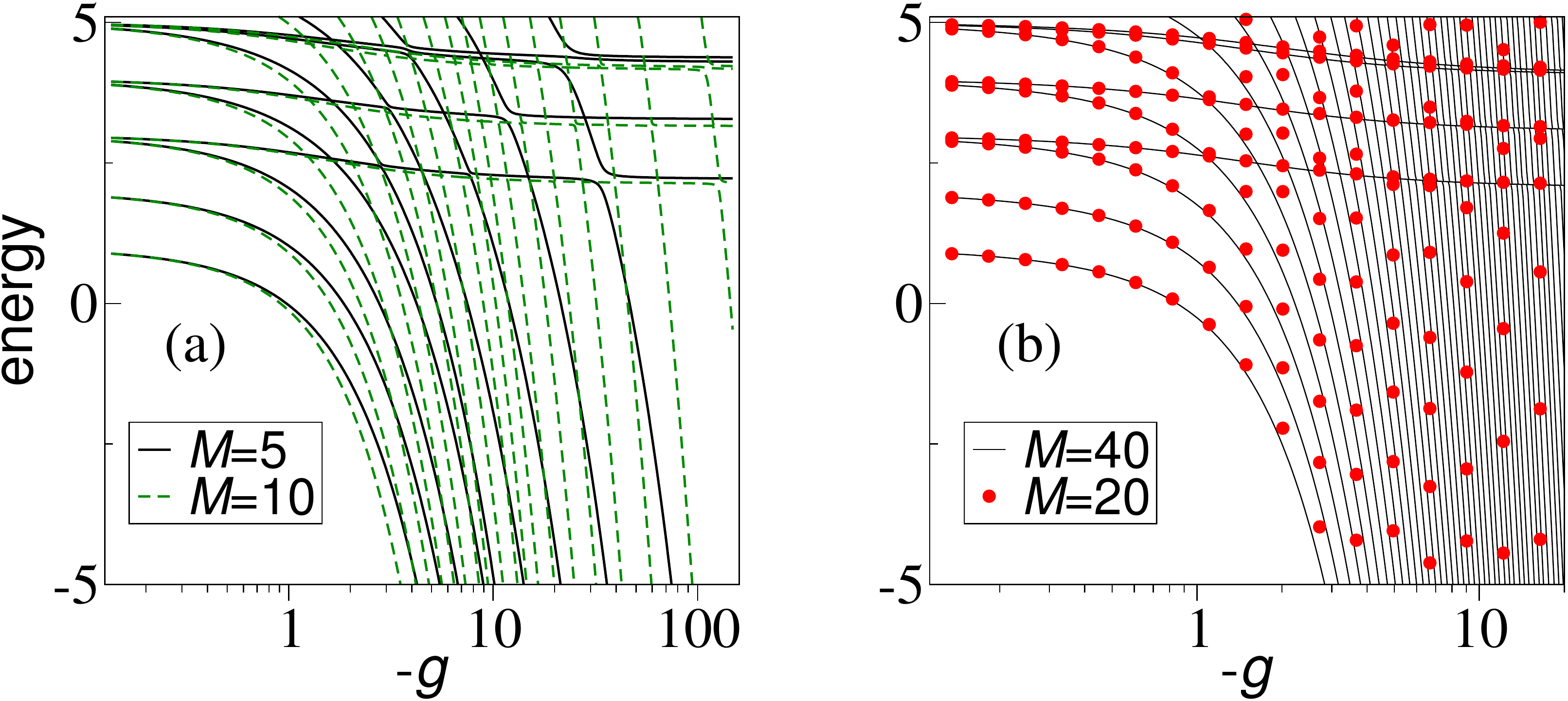}
\caption{ \label{fig:convergenceN2M20_M40}
%
% (Color online). 
%
Energy spectrum of the HO Hamiltonian 
\eqref{eq:LL2} for  $N=2$: role of the cutoff $M$. 
Already with a few  orbitals ($M=5$), the spectrum contains gas-like states 
with very weak dependence on interaction $g$ (note logarithmic scale).  The avoided level crossings
with non-horizontal states are prominent for small $M$.
Low lying gas like states, i.e. the sTG state, converge for relatively small $M$.  However cluster
states are connected to high energy states of the non-interacting system, so that the number and
energy of non-horizontal states is affected strongly by $M$.  }
\end{figure}
%%%%%%%%%%% FIGURE %%%%%%%%% FIGURE %%%%%%%%%%%%%% FIGURE %%%%%%%%% FIGURE

%%%%%%%%%%% FIGURE %%%%%%%%% FIGURE %%%%%%%%%%%%%% FIGURE %%%%%%%%% FIGURE
\begin{figure}[tb]
\centering
\includegraphics[width=0.98\columnwidth]{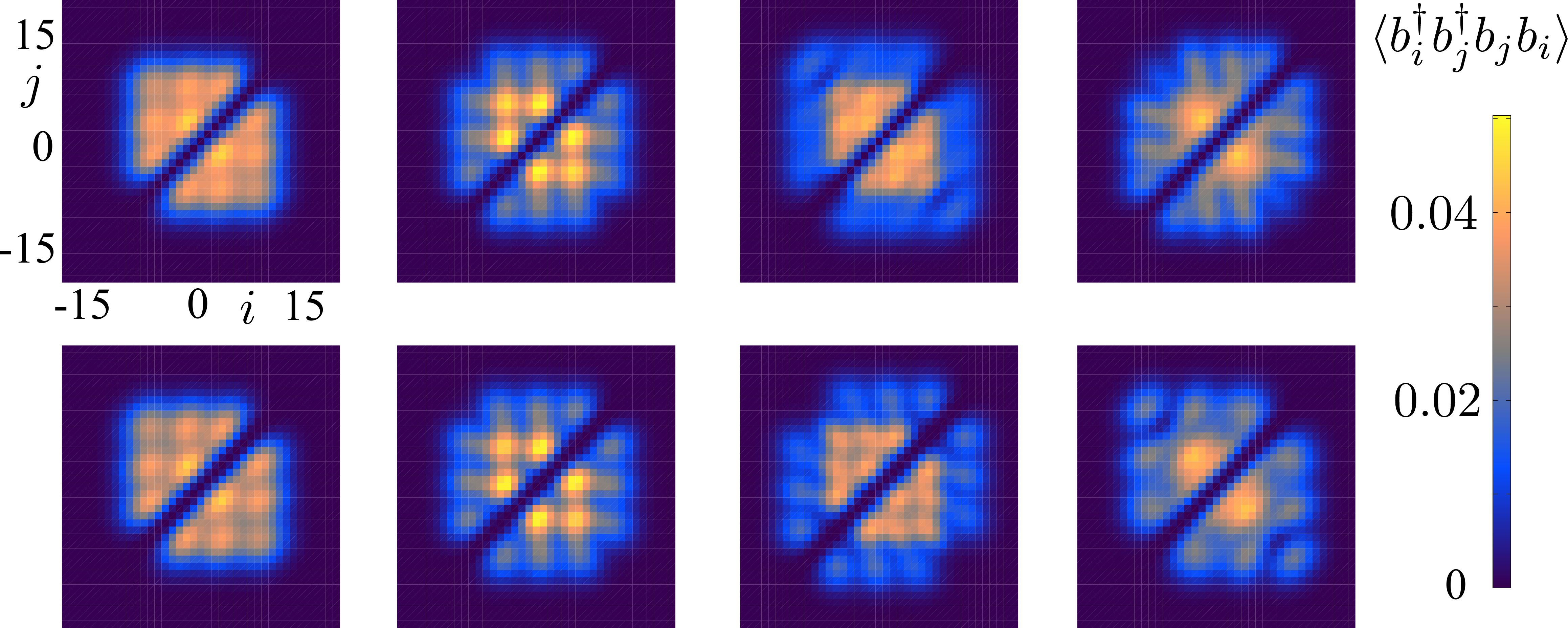}
\caption{ \label{fig:DenDenN4L40k0_005U20_02}
%
% (Color online). 
%
  Density-density correlation $\langle b_i^\dagger b_j^\dagger b_j b_i \rangle = \langle
  {n_i}{n_j}\rangle - n_i \delta_{ij}$ for $N=4, L=40, k=0.005$.  Upper panels show the four
  lowest-energy eigenstates for $U=20$.  Lower panels show the four lowest gas-like states for
  $U=-20$.}
\end{figure}
%%%%%%%%%%% FIGURE %%%%%%%%% FIGURE %%%%%%%%%%%%%% FIGURE %%%%%%%%%

The Lieb-Liniger Hamiltonian can be written in second quantization as
\begin{multline}
 H ~=~ \int dx\,\, \hat{\varPsi}^\dagger(x) \left( -\frac{\hbar^2\nabla^2}{2 m}
	  ~+~ \frac{1}{2}m\omega_0^2x^2 \right)\hat{\varPsi}(x) \\
     ~+~ g \,\int dx\,  \hat{\varPsi}^\dagger(x)\hat{\varPsi}^\dagger(x)	
		  \hat{\varPsi}(x)\hat{\varPsi}(x) \, .
\end{multline}
Since the single-particle harmonic-oscillator eigenstates $\varphi_k(x)$ form a complete eigenbasis,
the field operators can be expanded in the corresponding mode operators $a_k$:  
\begin{equation}
 \hat{\varPsi}(r) = \sum_k \varphi_k(r) a_k.
\end{equation}
The Hamiltonian then takes the form  \eqref{eq:LL2} with $M=\infty$ and 
\begin{equation}
f_{klmn} = \int dx\, \varphi_k^*(x) \varphi_l^*(x) \varphi_m(x) \varphi_n(x) 
\end{equation}
These integrals can be calculated conveniently using summation representations
\cite{Papenbrock_PRA2012, EdwardsEA_JRNIST96}.

Our approximation lies in using a finite number ($M$) of single-particle eigenstates.  The
$\tbinom{M+N-1}{N}$ basis states can then be chosen to be $\ket{n_1,n_2,...,n_M}$, where
$n_K\in\{0,1\}$ is the occupation of the $k$-th single-particle orbital and $\sum_kn_k=N$.

Fig.~\ref{fig:convergenceN2M20_M40} visualizes the influence of the cutoff $M$ on the energy
spectrum of two bosons, by comparing the spectra obtained using smaller and larger $M$ for
attractive interactions.  We see that the qualitative behavior of low-lying gas-like states is
captured already with small $M$.  However for small $M$ the sTG state has large avoided crossings
with cluster states; this gets corrected with increasing $M$.  There are strong corrections to the
cluster states as a function of $M$; the study of low energy cluster states requires large $M$ for
$g\ll-1$.

Thus, although we have an energy cutoff in the single-particle basis, the states best approximated
are not the lowest-energy many-particle states, but rather the lowest gas-like states.  This method
is thus particularly suitable for studying the sTG state and its dynamics.

\section{Density-density correlation for larger number of particles}

In the main text (Fig.~\ref{fig:denDenCorrN2L100U10_0_001}), we have described the structure of the
lowest gas-like states for $N=2$ in terms of density-density correlation functions.  The order of
the two states near $2\hbar \omega_0$ excitation, the breathing mode state and the 2nd Kohn-tower
state, were found to be interchanged between repulsive and attractive cases.

This situation is the same for larger number of particles, although the DDC plots are less
straightforward to interpret. In Fig.~\ref{fig:DenDenN4L40k0_005U20_02} we plot the DDC's for $N=4$.
For each $j$, there are generally three peaks as a function of $i$ (and vice versa), indicating the
probable positions of the other three particles.

The breathing mode eigenstate can be identified by the presence of a peak at large $j=-i$ and its mirror.  As
expected, this is the third eigenstate in the sequence for $U>0$, and the fourth for $U<0$.


\begin{thebibliography}{99}
  %Experiments


\bibitem{Cold-atom_reviews_expts} 
  M.~Lewenstein, A.~Sanpera, V.~Ahufinger, 
  \textit{Ultracold Atoms in Optical Lattices: Simulating quantum many-body systems}, 
  {Oxford University Press} (2012).
%
I.~Bloch, J.~Dalibard, and S.~Nascimb\'ene, Nature Physics {\bf 8,} 267  (2012).
%
I.~Bloch, J.~Dalibard, W.~Zwerger, Rev.\ Mod.\ Phys.\ {\bf 80}, 885 (2008).
%
T.~Kinoshita, T.~Wenger, and D.~S.~Weiss, Nature {\bf 440,} 900 (2006). 
%
M.~Greiner, O.~Mandel, T.~W.~H\"ansch, and I.~Bloch, Nature {\bf 419}, 51 (2002).


% \bibitem{cold_atom_expts} 

\bibitem{noneq_reviews}
%
A.~Polkovnikov, K.~Sengupta, A.~Silva, M.~Vengalattore,  Rev.\ Mod.\ Phys.\ {\bf 83}, 863 (2011). 
%
J.~Dziarmaga, Adv.\ Phys.\ {\bf 59}, 1063 (2010).


\bibitem{Winkler-etal_Nature06}
K.~Winkler, G.~Thalhammer, F.~Lang, R.~Grimm, J.~Hecker~Denschlag, A.~J.~Daley, A.~Kantian,
H.~P.~B\"uchler, and P.~Zoller, Nature  \textbf{441}, 853 (2006).  


\bibitem{bound_clusters_itinerant}
%
D.~Petrosyan, B.~Schmidt, J.~R.~Anglin, and M.~Fleischhauer, Phys. Rev. A \textbf{76}, 033606 (2007).
%
M.~Valiente and D.~Petrosyan, J.~Phys.\ B \textbf{41}, 161002 (2008).
%
L.~Jin, B.~Chen, and Z.~Song, Phys.\ Rev.\ A \textbf{79}, 032108 (2009). 
%
R.~A.~Pinto, M.~Haque, and S.~Flach, Phys.\ Rev.\ A {\bf 79}, 052118 (2009).
%
M.~Valiente, D.~Petrosyan, and A.~Saenz, Phys. Rev. A \textbf{81}, 011601(R) (2010).
%
R.~Khomeriki, D.~O.~Krimer, M.~Haque, and S.~Flach, Phys.\ Rev.\ A {\bf 81}, 065601 (2010).
%
A.~Deuchert, K.~Sakmann, A.~I.~Streltsov, O.~E.~Alon, and L.~S.~Cederbaum, Phys.\ Rev.\ A {\bf 86},
013618 (2012).
%
L.~F.~Santos and M.~I.~Dykman, New J.~Phys.\ \textbf{14}, 095019 (2012).



\bibitem{Naegerl_Science2009} E.~Haller, M.~Gustavsson, M.~J.~Mark, J.~G.~Danzl,  R.~Hart, G.~Pupillo, H.-C.~N\"agerl, Science \textbf{325}, 1224 (2009). 



\bibitem{KroenkeSchmelcher_PRA2013}
R.~Schmitz, S.~Kr\"onke, L.~Cao, and P.~Schmelcher, Phys.\ Rev.\ A \textbf{88}, 043601 (2013). 

\bibitem{our_PRA_2013} W.~Tschischik, R.~Moessner, M.~Haque, Phys.\ Rev.~A \textbf{88}, 063636 (2013).



\bibitem{AstrakharchikPRL} G.~E.~Astrakharchik, J.~Boronat, J.~Casulleras, and S.~Giorgini,
  Phys. Rev. Lett. \textbf{95}, 190407 (2005).
  
\bibitem{TempfliZollnerSchmelcher} E.~Tempfli, S.~ Z{\"o}llner, and P. Schmelcher, 
  New J. Phys. \textbf{10}, 103021 (2008).

\bibitem{Batchelor}M.~T.~Batchelor, M.~Bortz, X.~W.~Guan, and N.~Oelkers, J. 
  Stat. Mech.: Theory Exp. L10001 (2005).
  
\bibitem{ChenBetheAnsatz1} S.~Chen, L.~Guan, X.~Yin, Y.~Hao, and X.~Guan, Phys. Rev. A \textbf{81},
  031609(R) (2010).   
    
\bibitem{ChenBetheAnsatz2} Y.~Hao, H.~Guo, Y.~Zhang, and S.~Chen, Phys. Rev. A 
\textbf{83}, 053632 (2011).  

\bibitem{GirardeauAstrakharchikWaveFunctionssTG} M.~D.~Girardeau and G.~E.~Astrakharchik,
  Phys. Rev. A \textbf{81}, 061601(R) (2010).




\bibitem{KormosMussardoTrombettoni} M. Kormos, G. Mussardo, and A. 
  Trombettoni, Phys. Rev. A \textbf{83}, 013617 (2011).

\bibitem{Caux} M.~Panfil, J.~D.~Nardis, and J.-S.~Caux, Phys. Rev. Lett. 
  \textbf{110}, 125302 (2013).

\bibitem{ChenOverlapWithoutTrap} L.~Wang, Y.~Hao, and S.~Chen, Phys.
  Rev. A \textbf{81}, 063637 (2010).

\bibitem{Fleischhauer_TEBD} D.~Muth and M.~Fleischhauer, Phys. Rev. Lett.
  \textbf{105}, 150403 (2010).

\bibitem{Girardeau2} M. D. Girardeau, Phys. Rev. A \textbf{83}, 011601(R) 
  (2011).

% \bibitem{Busch_FoundPhys98} T.~Busch, B.-G. Engelert, K.~Rza\.{z}ewski, and M.~Wilkens,
%   Found. Phys. \textbf{28}, 549 (1998).

\bibitem{sTG_with_dipolar_bosons} M.~D.~Girardeau and G.~E.~Astrakharchik, Phys.\ Rev.\ Lett.\
  \textbf{109}, 235305 (2012). 
  

\bibitem{FermiGases} S.~Chen, X.~Guan, X.~Yin, L.~Guan, and M.~T.~Batchelor,
  Phys.\ Rev.~A \textbf{81}, 031608(R) (2010). M.~D.~Girardeau,  Phys.\ Rev.~A \textbf{82}, 011607(R)
  (2010).  L.~Guan and S.~Chen, Phys.\ Rev.\ Lett. \textbf{105}, 175301 (2010).  M.~D.~Girardeau,
  Phys. Rev. A \textbf{83}, 011601(R) (2011).  X.~Yin, X.~Guan, M.~T.~Batchelor, and S.~Chen,
  Phys. Rev. A \textbf{83}, 013602 (2011). 


\bibitem{LiebLiniger} E.~H.~Lieb and W.~Liniger, Phys. Rev. \textbf{130},  1605 (1963).


\bibitem{Olshanii_PRL1998} M.~Olshanii, Phys.\ Rev.\ Lett.\ \textbf{81}, 938 (1998).

\bibitem{Petrovetal_PRL2000} 
D.~S.~Petrov, G.~V.~Shlyapnikov, and J.~T.~M.~Walraven, Phys.\ Rev.\ Lett.\ \textbf{85}, 3745 (2000).



\bibitem{trapped_1D-boson_expts}
%
T.~Kinoshita, T.~Wenger, and D.~S.~Weiss, Science \textbf{305}, 1125 (2004).  Phys.\ Rev.\ Lett.\ \textbf{95},
190406 (2005); Nature (London) \textbf{440}, 900 (2006). 
%
A.~H.~van Amerongen, J.~J.~P.~van~Es, P.~Wicke, K.~V.~Kheruntsyan, and N.~J.~van Druten, 
Phys.\ Rev.\ Lett.\ \textbf{100}, 090402 (2008).
%
E.~Haller, R.~Hart, M.~J.~Mark, J.~G.~Danzl, L.~ Reichs{\"o}llner,  M.~Gustavsson, M.~Dalmonte,
G.~Pupillo, and H.-C.~N{\"a}gerl, Nature \textbf{466}, 597 (2010).  
%
T.~Jacqmin, J.~Armijo, T.~Berrada, K.~V.~Kheruntsyan, and I.~Bouchoule, Phys.\ Rev.\ Lett.\  \textbf{106}, 230405 (2011).
%
M.~J.~Davis, P.~B.~Blakie, A.~H.~van Amerongen, N.~J.~van Druten, and K.~V.~Kheruntsyan, Phys.\
Rev.~A  \textbf{85}, 031604(R)  (2012).  
%
A.~Vogler, R.~Labouvie, F.~Stubenrauch, G.~Barontini, V.~Guarrera, and H.~Ott, Phys.\ Rev.~A
\textbf{88}, 031603(R) (2013). 
%
B.~Fang, G.~Carleo, A.~Johnson, and I.~Bouchoule,   Phys.\ Rev.\ Lett.\  \textbf{113}, 035301 (2014).





\bibitem{HO_diagonalizations_various}
T.~Haugset and H.~Haugerud, Phys.\ Rev.\ A \textbf{57}, 3809 (1998).
%
T.~Papenbrock and G.~F.~Bertsch, Phys.\ Rev.~A \textbf{58}, 4854 (1998).
%
G.~F.~Bertsch and T.~Papenbrock, Phys.\ Rev.\ Lett.\ \textbf{83}, 5412 (1999). 
%
% T.~Papenbrock, Phys.\ Rev.~A \textbf{65}, 033606 (2002).
%
F.~Deuretzbacher, K.~Bongs, K.~Sengstock, and D.~Pfannkuche, Phys.\ Rev.~A \textbf{75}, 013614 (2007). 


\bibitem{Papenbrock_PRA2012}
T.~Papenbrock, Phys.\ Rev.~A \textbf{65}, 033606 (2002).


\bibitem{TGgasOpticalLattice} B.~Paredes, A.~Widera, V.~Murg, O.~Mandel, S.~F\"olling, I.~Cirac,
   G.~V.~Shlyapnikov, T.~W.~H\"ansch and I.~Bloch, Nature \textbf{429}, 277 (2004).


\bibitem{TEBD} G.~Vidal, PRL 91, 147902 (2003).
%
G.~Vidal, PRL 93, 040502 (2004).
%
M.~L.~Wall and L.~D.~Carr, Open Source TEBD,  \url{http://physics.mines.edu/downloads/software/tebd} (2009).   
  


\bibitem{KohnMode} L.~Brey, N.~F.~Johnson, and B.~I.~Halperin, Phys.\ Rev.\ B \textbf{40}, 10647
    (1989).  M.~Bonitz, K.~Balzer, and R.~van~Leeuwen, Phys. Rev. B \textbf{76}, 045341 (2007).


\bibitem{EdwardsEA_JRNIST96} M.~Edwards, R.~J.~Dodd, C.~W.~Clark, K.~Burnett, J.~Res.\ Nat.\ Insit.\ St.\
  Tech. \textbf{101}, 553 (1996). 


\bibitem{Bonitz_breathing_mode_review}
J.~W.~Abraham and M.~Bonitz, Contributions to Plasma Physics, \textbf{54}, 27 (2014).  




\end{thebibliography}
\end{document}